\documentclass{amsart}
\usepackage{amssymb}
\usepackage{amsmath}
\usepackage{amsfonts}

\newtheorem{theorem}{Theorem}
\theoremstyle{plain}

\newtheorem{claim}[theorem]{Claim}

\newtheorem{corollary}[theorem]{Corollary}

\newtheorem{example}[theorem]{Example}

\newtheorem{lemma}[theorem]{Lemma}

\newtheorem{problem}[theorem]{Problem}
\newtheorem{proposition}[theorem]{Proposition}
\newtheorem{remark}[theorem]{Remark}

\numberwithin{equation}{section}
\numberwithin{theorem}{section}
\input{tcilatex}

\begin{document}
\title[Diagonalization of the discrete Fourier transform]{On the
diagonalization of the discrete Fourier transform}
\author{Shamgar Gurevich}
\address{Department of Mathematics, University of California, Berkeley, CA
94720, USA. }
\email{shamgar@math.berkeley.edu}
\author{Ronny Hadani}
\address{Department of Mathematics, University of Chicago, IL, 60637, USA.}
\email{hadani@math.uchicago.edu}
\date{Dec. 1, 2007. }
\thanks{\copyright \ Copyright by S. Gurevich and R. Hadani, Dec. 1, 2007.
All rights reserved.}
\dedicatory{Dedicated to William Kahan and Beresford Parlett on the occasion
of their 75th birthday}

\begin{abstract}
The discrete Fourier transform (DFT) is an important operator which acts on
the Hilbert space of complex valued functions on the ring $%
\mathbb{Z}
/N%
\mathbb{Z}
$. In the case where $N=p$ is an odd prime number, we exhibit a canonical
basis $\Phi $ of eigenvectors for the DFT. The transition matrix $\Theta $
from the standard basis to $\Phi $ defines a novel transform which we call
the \textit{discrete oscillator transform }(DOT for short).\textit{\ }%
Finally, we describe a fast algorithm for computing $\Theta $ in certain
cases.
\end{abstract}

\maketitle

\section{Introduction}

The discrete Fourier transform (DFT) is probably one of the most important
operators in modern science. It is omnipresent in various fields of discrete
mathematics and engineering, including combinatorics, number theory,
computer science and, last but probably not least, digital signal
processing. Formally, the DFT is a family $\left \{ F_{N}\right \} $ of \
unitary operators, where each $F_{N}$ acts on the Hilbert space $\mathcal{H}%
_{N}\mathcal{=%
\mathbb{C}
}\left( 
\mathbb{Z}
/N%
\mathbb{Z}
\right) $ by the formula 
\begin{equation*}
F_{N}\left[ f\right] \left( w\right) =\frac{1}{\sqrt{N}}\sum \limits_{t\in 
\mathbb{Z}
/N%
\mathbb{Z}
}e^{\frac{2\pi i}{N}wt}f\left( t\right) .
\end{equation*}

Although so widely used, the spectral properties of the DFT remains to some
extent still mysterious. For example, the calculation of the multiplicities
of its eigenvalues, which was first carried out by Gauss, is quite involved
and requires a multiple of number theoretic manipulations \cite{AT}.

A primary motivation for studying the eigenvectors of the DFT comes from
digital signal processing. Here, a function is considered in two basic
realizations: The time realization and the frequency realization. Each
realization, yields information on different attributes of the function. The
DFT operator acts as a dictionary between these two realizations%
\begin{equation*}
\text{\textbf{Time} }\overset{F_{N}}{\rightleftarrows }\text{ \textbf{%
Frequency.} }
\end{equation*}

From this point of view, it is natural to look for a diagonalization basis,
namely, a basis of eigenvectors (eigen modes) for $F_{N}$. In this regard,
the main conceptual difficulty comes from the fact that the diagonalization
problem is ill-defined, since, $F_{N}$ is an operator of order 4, i.e., $%
F_{N}^{4}=Id$, which means that it has at most four eigenvalues $\pm 1$,$\pm
i$, therefore each appears with large multiplicity (assuming $N\gg 4$).

An interesting approach to the resolution of this difficulty, motivated from
results in continuous Fourier analysis, was developed by Gr\"{u}nbaum in 
\cite{G}. In that approach, a tridiagonal operator $S_{N}$ which commutes
with $F_{N}$ and admits a simple spectrum is introduced. This enable him to
give a basis of eigenfunctions for the DFT. Specifically, $S_{N}$ appears as
a certain discrete analogue of the differential operator $D=\partial
_{t}^{2}-t^{2}$ which commutes with the continuous Fourier transform. Other
approaches can be found in \cite{DS, M}.

\subsection{Main results of this paper}

In this paper we describe a representation theoretic approach to the
diagonalization problem of the DFT in the case when $N=p$ is an odd prime
number. Our approach, puts to the forefront the Weil representation \cite{W}
of the finite symplectic group $Sp=SL_{2}\left( \mathbb{F}_{p}\right) $ as
the fundamental object underlying harmonic analysis in the finite setting
(see \cite{FHKMN, GHH, GHS1, GHS2, GHS3, V} for other recent applications of
this point of view).

Specifically, we exhibit a canonical basis $\Phi _{p}$ of eigenvectors for
the DFT. We also describe the transition matrix $\Theta _{p}$ from the
standard basis to $\Phi _{p}$, which we call the \textit{discrete oscillator
transform} (DOT for short). In addition, in the case $p\equiv 1\left( \func{%
mod}4\right) $, we describe a fast algorithm for computing $\Theta _{p}$
(FOT for short). More precisely, we explain how in this case one can reduce
the DOT to a composition of transforms with existing fast algorithms.

It is our general feeling that the Weil representation yields a transparent
explanation to many classical results in finite harmonic analysis. In
particular, we describe an alternative method for calculating the
multiplicities of the eigenvalues for the DFT, a method we believe is more
suggestive then the classical calculations.

The rest of the introduction is devoted to a more detailed account of the
main ideas and results of this paper.

\subsection{Symmetries of the DFT}

Let us fix an odd prime number $p$ and for the rest of the introduction
suppress the subscript $p$ from all notations.

Generally, when a (diagonalizable) linear operator $A$ has eigenvalues
admitting large multiplicities, it suggest that there exists a group $%
G=G_{A}\subset GL\left( \mathcal{H}\right) $ of "hidden" symmetries
consisting of operators which commute with $A$. \ Alas, usually the problem
of computing the group $G$ is formidable and, in fact, equivalent to the
problem of diagonalizing $A$. If the operator $A$ \ arises "naturally",
there is a chance that the group $G$ can be effectively described. In
preferred situations, $G$ is commutative and large enough so that all
degeneracies are resolved and the spaces of common eigenvectors with respect
to $G$ are one-dimensional. The basis of common eigenvectors with respect to 
$G$ establishes a distinguish choice of eigenvectors for $A$.
Philosophically, we can say that it is more correct to consider from start
the group $G$ instead of the single operator $A$.

Interestingly, the DFT operator $F$ admits a natural group of symmetries $%
G_{F}$, which, in addition, can be effectively described using the Weil
representation. For the sake of the introduction, it is enough to know that
the Weil representation in this setting is a unitary representation $\rho
:Sp\rightarrow U\left( \mathcal{H}\right) $ and the key observation is that $%
F$ is proportional to a single operator $\rho \left( \mathrm{w}\right) $.
The group $G_{F}$ is the image under $\rho $ of the centralizer subgroup $T_{%
\mathrm{w}}$ of $\mathrm{w}$ in $Sp.$

\subsection{The algebraic torus associated with the DFT}

The subgroup $T_{\mathrm{w}}$ can be computed explicitly and is of a very
"nice" type, it consists of rational points of a maximal algebraic torus in $%
Sp$ which concretely means that it is maximal commutative subgroup in $Sp$,
consisting of elements which are diagonalizable over some field extension.
Restricting the Weil representation to the subgroup $T_{\mathrm{w}}$ yields
a collection $G_{F}=\left \{ \rho \left( g\right) :g\in T_{\mathrm{w}%
}\right
\} $ of commuting operators, each acts unitarily on the Hilbert
space $\mathcal{H}$ and commutes with $F$. This, in turn, yields a
decomposition, stable under Fourier transform, into character spaces 
\begin{equation}
\mathcal{H=}\bigoplus \mathcal{H}_{\chi },  \label{dec1_eq}
\end{equation}%
where $\chi $ runs in the set of (complex valued) characters of $T_{\mathrm{w%
}}$, namely, $\phi \in \mathcal{H}_{\chi }$ iff $\rho \left( g\right) \phi
=\chi \left( g\right) \phi $ for every $g\in T_{\mathrm{w}}$. The main
technical statement of this paper, Theorem \ref{dec_thm}, roughly says that $%
\dim \mathcal{H}_{\chi }=1$ for every $\chi $ which appears in (\ref{dec1_eq}%
).

\subsection{The oscillator transform}

Choosing a unit representative $\phi _{\chi }\in \mathcal{H}_{\chi }$ for
every $\chi $, gives the canonical basis $\Phi =\left \{ \phi _{\chi
}\right
\} $ of eigenvectors for $F$. The oscillator transform $\Theta $
sends a function $f\in \mathcal{H}$ to the coefficients in the unique
expansion 
\begin{equation*}
f=\sum a_{\chi }\phi _{\chi }.
\end{equation*}

The fine behavior of $F$ and $\Theta $ is governed by the\ (split type)
structure of $T_{\mathrm{w}}$, which changes depending on the value of the
prime $p$ modulo $4$. This phenomena has several consequences. In
particular, it gives a transparent explanation to the precise way the
multiplicities of the eigenvalues of $F$ depend on the prime $p$. Another,
algorithmic, consequence is related to the existence of a fast algorithm for
computing $\Theta $.

\subsection{Remarks}

\subsubsection{Field extension}

All the results in this paper were stated for the basic finite field $%
\mathbb{F}_{p}$, for the reason of making the terminology more accessible.
In fact, all the results can be stated and proved in essentially the same
way for any field extension $\mathbb{F}_{q}$, $q=p^{n}$. One just need to
replace $p$ by $q$ in all appropriate places. The only non-trivial statement
in this respect is Corollary \ref{Fq}.

\subsubsection{Properties of eigenvectors}

The character vectors $\phi _{\chi }$ satisfy many interesting properties
and are objects of study in their own right. A comprehensive treatment of
this aspect of the theory appears in \cite{GHS1, GHS3}.

\subsection{Structure of the paper}

The paper consists of two main parts except of the introduction:

\begin{itemize}
\item In Section \ref{pre_sec}, we describe the Weil representation. We
begin by discussing the finite Heisenberg group and the Heisenberg
representation. Then, we introduce the Weil representation of the finite
symplectic group, first it is described in abstract terms and then more
explicitly invoking the idea of invariant presentation of an operator. In
Section \ref{Tori}, we discuss the theory of tori in the one-dimensional
Weil representation, and in Section \ref{OT}, we explain how to associate to
a maximal torus $T\subset Sp$ a transform $\Theta _{T}$ that we call the
discrete oscillator transform. In addition, we describe a fast algorithm for
computing $\Theta _{T}$ in the case $T$ is a split torus. Finally, in
Section \ref{Diag}, we apply the theory to the specific torus associated
with the DFT operator. We finish this section with a treatment of the
multiplicity problem for the DFT, from the representation theoretic
perspective.

\item We have two Appendices. Appendix \ref{proofs_appendix}, consisting of
proofs of statements which appeared in Sections \ref{pre_sec}-\ref{Diag},
and Appendix \ref{EOT}, consisting of an explicit formula for the particular
oscillator transform $\Theta _{T_{\mathrm{w}}}$ associated with the DFT.
\end{itemize}

\subsection{Acknowledgements}

It is a pleasure to thank our teacher J. Bernstein for his interest and
guidance. We acknowledge R. Howe for sharing with us some of his thoughts
concerning the finite Weil representation. We are happy to thank A. Sahai
and N. Sochen for all of our discussions on the applied aspects of this work.

We appreciate several discussions we had with A. Gr\"{u}nbaum, W. Kahan, B.
Parlett and B. Porat on the DFT. We thank \ A. Weinstein, J. Wolf \ and M.
Zworski for interesting conversations related to the Weil representation.
Finally, we thank P. Diaconis, M. Gu, M. Haiman, B. Poonen and K. Ribet for
the opportunities to present this work in the MSRI, number theory, RTG and
scientific computing seminars at Berkeley during February 2008.

\section{The Weil representation\label{pre_sec}}

\subsection{The Heisenberg group}

Let $(V,\omega )$ be a two-dimensional symplectic vector space over the
finite field $\mathbb{F}_{p}$. The reader should think of $V$ as $\mathbb{F}%
_{p}\times \mathbb{F}_{p}$ with the standard form $\omega \left( \left(
t,w\right) ,\left( t^{\prime },w^{\prime }\right) \right) =tw^{\prime
}-wt^{\prime }$. Considering $V$ as an abelian group, it admits a
non-trivial central extension called the \textit{Heisenberg }group. \
Concretely, the group $H$ can be presented as the set $H=V\times \mathbb{F}%
_{p}$ with the multiplication given by%
\begin{equation*}
(v,z)\cdot (v^{\prime },z^{\prime })=(v+v^{\prime },z+z^{\prime }+\tfrac{1}{2%
}\omega (v,v^{\prime })).
\end{equation*}

The center of $H$ is $\ Z=Z(H)=\left \{ (0,z):\text{ }z\in \mathbb{F}%
_{p}\right \} .$ The symplectic group $Sp=Sp(V,\omega )$, which in this case
is isomorphic to $SL_{2}\left( \mathbb{F}_{p}\right) $, acts by automorphism
of $H$ through its action on the $V$-coordinate.

\subsection{The Heisenberg representation\label{HR}}

One of the most important attributes of the group $H$ is that it admits a
special family of irreducible representations. The precise statement goes as
follows. Let $\psi :Z\rightarrow 
\mathbb{C}
^{\times }$ be a non-trivial character of the center. For example in this
paper we take $\psi \left( z\right) =e^{\frac{2\pi i}{p}z}$. It is not hard
to show

\begin{theorem}[Stone-von Neuman]
\label{S-vN}There exists a unique (up to isomorphism) irreducible unitary
representation $(\pi ,H,\mathcal{H)}$ with the center acting by $\psi ,$
i.e., $\pi _{|Z}=\psi \cdot Id_{\mathcal{H}}$.
\end{theorem}

The representation $\pi $ which appears in the above theorem will be called
the \textit{Heisenberg representation}.

\subsubsection{Standard realization of the Heisenberg representation\label%
{standard_subsub}.}

The Heisenberg representation $(\pi ,H,\mathcal{H)}$ can be realized as
follows: $\mathcal{H}$ is the Hilbert space $%
\mathbb{C}
(\mathbb{F}_{p})$ of complex valued functions on the finite line, with the
standard Hermitian product. The action $\pi $ is given by

\begin{itemize}
\item $\pi (t,0)[f]\left( x\right) =f\left( x+t\right) $;

\item $\pi (0,w)[f]\left( x\right) =\psi \left( wx\right) f\left( x\right) $;

\item $\pi (z)[f]\left( x\right) =\psi \left( z\right) f\left( x\right) ,$
\end{itemize}

for every $f\in \mathcal{H}$, $x,t,w\in \mathbb{F}_{p}$ and $z\in Z.$

We call this explicit realization the \textit{standard realization}.

\subsection{The Weil representation\label{Wrep_sub}}

A direct consequence of Theorem \ref{S-vN} is the existence of a projective
representation $\widetilde{\rho }:Sp\rightarrow PGL(\mathcal{H)}$. The
construction of $\widetilde{\rho }$ out of the Heisenberg representation $%
\pi $ is due to Weil \cite{W} and it goes as follows. Considering the
Heisenberg representation $\pi $ and an element $g\in Sp$, one can define a
new representation $\pi ^{g}$ acting on the same Hilbert space via $\pi
^{g}\left( h\right) =\pi \left( g\left( h\right) \right) $. Clearly, both $%
\pi $ and $\pi ^{g}$ have the same central character $\psi ,$ hence, by
Theorem \ref{S-vN}, they are isomorphic. Since the space $\mathsf{Hom}%
_{H}(\pi ,\pi ^{g})$ is one-dimensional, choosing for every $g\in Sp$ a
non-zero representative $\widetilde{\rho }(g)\in \mathsf{Hom}_{H}(\pi ,\pi
^{g})$ gives the required projective Weil representation. In more concrete
terms, the projective representation $\widetilde{\rho }$ is characterized by
the formula 
\begin{equation}
\widetilde{\rho }\left( g\right) \pi \left( h\right) \widetilde{\rho }\left(
g^{-1}\right) =\pi \left( g\left( h\right) \right) ,  \label{Egorov}
\end{equation}%
for every $g\in Sp$ and $h\in H$. \ A more delicate statement is that there
exists a unique lifting of $\widetilde{\rho }$ into a linear representation.

\begin{theorem}
\label{linearization_thm} The projective Weil representation uniquely%
\footnote{%
Uniquely, except in the case the finite field is $\mathbb{F}_{3}$. For the
canonical choice in the latter case see \cite{GH1}.} lifts to a linear
representation 
\begin{equation*}
\rho :SL_{2}(\mathbb{F}_{p})\longrightarrow GL(\mathcal{H)},
\end{equation*}

that satisfies equation (\ref{Egorov}).
\end{theorem}

The existence of a linearization $\rho $ follows from a known fact \cite{B}
that any\footnote{%
A more direct proof of the linearity of the Weil representation exists in 
\cite{GH1, GH2}.} projective representation of $SL_{2}(\mathbb{\mathbb{F}}%
_{p})$ can be linearized to an honest representation. The uniqueness follows
from the well known fact (we give an argument in Appendix \ref%
{proofs_appendix}) that the group $SL_{2}(\mathbb{F}_{p})$, $p\neq 3$, is
perfect, i.e., it has no non-trivial characters.

\subsubsection{Invariant presentation of the Weil representation}

Let us denote by $%
\mathbb{C}
\left( H,\psi \right) $ the space of (complex valued) functions on $H$ which
are $\psi $-equivariant with respect to the action of the center, namely, a
function $f\in 
\mathbb{C}
\left( H,\psi \right) $ satisfies $f\left( zh\right) =\psi \left( z\right)
f\left( h\right) $ for every $z\in Z$, $h\in H$. Given an operator $A\in 
\mathsf{End}\left( \mathcal{H}\right) $, it can be written in a unique way
as $A=\pi \left( K_{A}\right) $, where $K_{A}\in 
\mathbb{C}
\left( H,\psi ^{-1}\right) $ and $\pi $ denotes the extended action $\pi
\left( K_{A}\right) =\sum \limits_{h\in H}K_{A}\left( h\right) \pi \left(
h\right) .$ The function $K_{A}$ is called the \textit{kernel of }$A$ and it
is given by the \textit{matrix coefficient }%
\begin{equation}
K_{A}\left( h\right) =\tfrac{1}{\dim \mathcal{H}}Tr\left( A\pi \left(
h^{-1}\right) \right) .  \label{Weyl_eq}
\end{equation}

In the context of the Heisenberg representation, formula (\ref{Weyl_eq}) is
usually referred to as the \textit{Weyl transform \cite{We}}.

Using the Weyl transform we are able to give an explicit description of the
Weil representation. The idea \cite{GH1} is to write each operator $\rho
\left( g\right) $, $g\in Sp,$ in terms of its kernel function $K_{g}=K_{\rho
\left( g\right) }\in 
\mathbb{C}
\left( H,\psi ^{-1}\right) $. The following formula is taken from \cite{GH1}%
\begin{equation}
K_{g}\left( v,z\right) =\tfrac{1}{\dim \mathcal{H}}\sigma \left( -\det
\left( \kappa \left( g\right) +I\right) \right) \cdot \psi \left( \tfrac{1}{4%
}\omega \left( \kappa \left( g\right) v,v\right) +z\right)
\label{inv_formula}
\end{equation}%
for every $g\in Sp$ such that $g-I$ is invertible, where $\sigma $ denotes
the unique quadratic character (Legendre character) of the multiplicative
group $\mathbb{F}_{p}^{\times }$ and $\kappa \left( g\right) =\frac{g+I}{g-I}
$ is the Cayley transform.

\begin{remark}
Sketch of the proof of (\ref{inv_formula}) (see details in \cite{GH1}).
First, one can easily show that 
\begin{equation*}
K_{g}\left( v,z\right) =\tfrac{1}{\dim \mathcal{H}}\mu _{g}\cdot \psi \left( 
\tfrac{1}{4}\omega \left( \kappa \left( g\right) v,v\right) +z\right) ,
\end{equation*}%
for $g\in $ $Sp$ with $g-I$ invertible, and for some $\mu _{g}\in 
\mathbb{C}
$. \ Second, for $\rho $ to be a linear representation the kernel $%
K_{g}:H\rightarrow 
\mathbb{C}
$ should satisfy the multiplicativity property%
\begin{equation}
K_{g_{1}\cdot g_{2}}=K_{g_{1}}\ast K_{g_{2}},  \label{conv_eq}
\end{equation}%
where the $\ast $ operation denotes convolution with respect to the
Heisenberg group action. Finally, a direct calculation reveals that $\mu
_{g} $ must be equal to $\sigma \left( -\det \left( \kappa \left( g\right)
+I\right) \right) $ for (\ref{conv_eq}) to hold. Hence, we obtain (\ref%
{inv_formula}).
\end{remark}

\section{The theory of tori in the Weil representation\label{Tori}}

\subsection{Tori}

A maximal (algebraic) torus in $Sp$ is a maximal commutative subgroup which
becomes diagonalizable over some field extension. There exists two conjugacy
classes of maximal (algebraic) tori in $Sp$. The first class consists of
those tori which are diagonalizable already over $\mathbb{F}_{p}$ or
equivalently those are the tori that are conjugated to the standard diagonal
torus%
\begin{equation*}
A=\left \{ 
\begin{pmatrix}
a & 0 \\ 
0 & a^{-1}%
\end{pmatrix}%
:a\in \mathbb{F}_{p}\right \} .
\end{equation*}

A torus in this class is called a \textit{split} torus. The second class
consists of those tori which become diagonalizable over a quadratic
extension $\mathbb{F}_{p^{2}}$ or equivalently those are tori which are not
conjugated to $A.$ A torus in this class is called a \textit{non-split }%
torus (sometimes it is called inert torus)$.$

\begin{example}[Example of a non-split torus]
It might be suggestive to explain further the notion of non-split torus by
exploring, first, the analogue notion in the more familiar setting of the
field $%
\mathbb{R}
$. Here, the standard example of a maximal non-split torus is the circle
group $SO(2)\subset SL_{2}(%
\mathbb{R}
)$. Indeed, it is a maximal commutative subgroup which becomes
diagonalizable when considered over the extension field $%
\mathbb{C}
$ of complex numbers. The above analogy suggests a way to construct an
example of a maximal non-split torus in the finite field setting as well.

Let us identify the symplectic plane $V=\mathbb{F}_{p}\times \mathbb{F}_{p}$
with the quadratic extension $\mathbb{F}_{p^{2}}$. Under this
identification, $\mathbb{F}_{p^{2}}$ acts on $V$ and for every $g\in $ $%
\mathbb{F}_{p^{2}}$ we have $\omega \left( gu,gv\right) =\det \left(
g\right) \omega \left( u,v\right) $, which implies that the group 
\begin{equation*}
T_{ns}=\left \{ g\in \mathbb{F}_{p^{2}}^{\times }:\det \left( g\right)
=1\right \}
\end{equation*}%
naturally lies in $Sp$. The group $T_{ns}$ is an example of a non-split
torus which the reader might think of as the\ "finite circle".
\end{example}

\subsection{Decompositions with respect to a maximal torus}

Restricting the Weil representation to a maximal torus $T\subset Sp$ yields
a decomposition 
\begin{equation}
\mathcal{H=}\tbigoplus_{\chi }\mathcal{H}_{\chi },  \label{decomp_eq}
\end{equation}%
where $\chi $ runs in the set $T^{\vee }$ of complex valued characters of
the torus $T$. More concretely, choosing a generator\footnote{%
A maximal torus $T$ in $SL_{2}\left( \mathbb{F}_{p}\right) $ is a cyclic
group, thus there exists a generator.} $t\in T$, the decomposition (\ref%
{decomp_eq}) naturally corresponds to the eigenspaces of the linear operator 
$\rho \left( t\right) $. The decomposition (\ref{decomp_eq}) depends on the
split type of $T$. Let $\sigma _{T}$ denote the unique quadratic character
of $T$.

\begin{theorem}
\label{dec_thm}If $T$ is a split torus then
\end{theorem}

\begin{equation*}
\dim \mathcal{H}_{\chi }=\left \{ 
\begin{array}{cc}
1 & \chi \neq \sigma _{T}, \\ 
2 & \chi =\sigma _{T}.%
\end{array}%
\right.
\end{equation*}%
If $T$ is a non-split torus then 
\begin{equation*}
\dim \mathcal{H}_{\chi }=\left \{ 
\begin{array}{cc}
1 & \chi \neq \sigma _{T}, \\ 
0 & \chi =\sigma _{T}.%
\end{array}%
\right.
\end{equation*}

For a proof see Appendix \ref{proofs_appendix}.

\section{ The oscillator transform associated with a maximal torus\label{OT}}

Let us fix a maximal torus $T\subset Sp$. Every vector $f\in \mathcal{H}$
can be written uniquely as a direct sum $f=\sum f_{\chi }$ with $f_{\chi
}\in \mathcal{H}_{\chi }$ and $\chi $ runs in $I=\mathsf{Spec}_{T}\left( 
\mathcal{H}\right) $ - the spectral support of $\mathcal{H}$ with respect to 
$T$ consisting of all characters $\chi \in T^{\vee }$ such that $\dim 
\mathcal{H}_{\chi }\neq 0$. Let us choose, in addition, a collection of unit
vectors $\phi _{\chi }\in \mathcal{H}_{\chi }$, $\chi \in I,$ and let $\phi
= $ $\sum \phi _{\chi }$. We define the transform $\Theta _{T}=\Theta
_{T,\phi }:\mathcal{H\rightarrow 
\mathbb{C}
}\left( I\right) $ by 
\begin{equation}
\Theta _{T}\left[ f\right] \left( \chi \right) =\left \langle f,\phi _{\chi
}\right \rangle .  \label{Theta-T}
\end{equation}%
We will call the transform $\Theta _{T}$ the \textit{discrete} \textit{%
oscillator transform} (DOT for short) with respect to the torus $T$ and the
test vector $\phi $.

\begin{remark}
We note that in the case $T$ is a non-split torus, $\Theta _{T}$ maps $%
\mathcal{H}$ isomorphically to $\mathcal{%
\mathbb{C}
}\left( I\right) $. In the case $T$ is a split torus, $\Theta _{T}$ has a
kernel consisting of $f\in \mathcal{H}$ such that $\left \langle f,\phi
_{\sigma _{T}}\right \rangle =0$.
\end{remark}

\subsection{The oscillator transform (integral form)}

Let $\mathcal{M}_{T}:%
\mathbb{C}
(T)\rightarrow 
\mathbb{C}
(T^{\vee })$ denote the Mellin transform 
\begin{equation*}
\mathcal{M}_{T}\left[ f\right] \left( \chi \right) =\frac{1}{\#T}\sum
\limits_{g\in T}\overline{\chi }\left( g\right) f\left( g\right) ,
\end{equation*}%
for $f\in 
\mathbb{C}
\left( T\right) $, where $\overline{\chi }$ denotes the complex conjugate of 
$\chi .$ Let us denote by $m_{T}:\mathcal{H\rightarrow 
\mathbb{C}
}\left( T\right) $ the matrix coefficient $m_{T}\left[ f\right] \left(
g\right) =\left \langle f,\rho \left( g^{-1}\right) \phi \right \rangle $
for $f\in \mathcal{H}$.

\begin{lemma}
\label{integral_lemma}We have%
\begin{equation*}
\Theta _{T}=\mathcal{M}_{T}\circ m_{T}.
\end{equation*}
\end{lemma}

For a proof see Appendix \ref{proofs_appendix}.

\subsection{Fast oscillator transforms\label{fast_subsub}}

In practice, it is desirable to have a "fast" algorithm for computing the
oscillator transform (FOT for short). We work in the following setting. The
vector $f$ is considered in the standard realization $\mathcal{H=%
\mathbb{C}
}\left( \mathbb{F}_{p}\right) $ (see \ref{standard_subsub}). In this context
the oscillator transform gives the transition matrix between the basis of
delta functions and the basis $\Phi _{T}=\left \{ \phi _{\chi }\right \} $
of character vectors. We will show that when $T$ \ is a split torus and for
an appropriate choice of $\phi $, the oscillator transform can be computed
in $O\left( p\log \left( p\right) \right) $ arithmetic operations.
Principally, what we will show is that the computation reduces to an
application of DFT followed by an application of the standard Mellin
transform, both transforms admit a fast algorithm \cite{CT}.

Assume $T$ is a split torus. Since all split tori are conjugated to one
another, there exists, in particular, an element $s\in Sp$ conjugating $T$
with the standard diagonal torus $A$. In more details, we have a
homomorphism of groups $Ad_{s}:T\rightarrow A$ sending $g\in T$ to $%
Ad_{s}\left( g\right) =sgs^{-1}\in A$. Dually, we have a homomorphism $%
Ad_{s}^{\vee }:A^{\vee }\rightarrow T^{\vee }$ between the corresponding
groups of characters.

The main idea is to relate the oscillator transform with respect to $T$ with
the oscillator transform with respect to $A$. The relation is specified in
the following simple lemma.

\begin{lemma}
\label{relation_lemma}We have%
\begin{equation}
\left( Ad_{s}^{\vee }\right) ^{\ast }\circ \Theta _{T,\phi }=\Theta _{A,\rho
\left( s\right) \phi }\circ \rho \left( s\right) .  \label{relation_eq}
\end{equation}
\end{lemma}

For the proof see Appendix \ref{proofs_appendix}.

\begin{remark}
Roughly speaking, (\ref{relation_eq}) means that (up to a
"reparametrization" of $T^{\vee }$ by $A^{\vee }$ using $Ad_{s}^{\vee }$)
the oscillator transform$\ $of a vector $f\in \mathcal{H}$ with respect to
the torus $T$ is the same as the oscillator transform of the vector $\rho
\left( s\right) f$ $\ $with respect to the diagonal torus $A$.
\end{remark}

In order to finish the construction of the fast algorithm we need to recall
some well-known facts about the Weil representation in the standard
realization.

First, we recall that the group $Sp$ admits a Bruhat decomposition $Sp=B\cup
B\mathrm{w}B,$ where $B$ denote the Borel subgroup of lower triangular
matrices and $\mathrm{w}$ denotes the Weyl element%
\begin{equation}
\mathrm{w}=%
\begin{pmatrix}
0 & 1 \\ 
-1 & 0%
\end{pmatrix}%
.  \label{Weyl}
\end{equation}%
Furthermore, the Borel subgroup $B$ can be written as a product $B=AU=UA$,
where $A$ is the standard diagonal torus and $U$ is the unipotent subgroup 
\begin{equation*}
U=\left \{ 
\begin{pmatrix}
1 & 0 \\ 
b & 1%
\end{pmatrix}%
:b\in \mathbb{F}_{p}\right \} .
\end{equation*}

Therefore, we can write the Bruhat decomposition also as $Sp=UA\cup U\mathrm{%
w}UA$.

Second, we give an explicit description of the operators in the Weil
representation, associated with different types of elements in $Sp$. The
operators are specified up to a unitary scalar:

\begin{itemize}
\item The standard torus $A$ acts by (normalized) scaling: An element $g=%
\begin{pmatrix}
a & 0 \\ 
0 & a^{-1}%
\end{pmatrix}%
,$ acts by 
\begin{equation*}
S_{a}\left[ f\right] \left( x\right) =\sigma \left( a\right) f\left(
a^{-1}x\right) .
\end{equation*}

\item The group $U$ of unipotent matrices acts by quadratic characters
(chirps): An element $g=%
\begin{pmatrix}
1 & 0 \\ 
b & 1%
\end{pmatrix}%
$, acts by 
\begin{equation*}
M_{b}\left[ f\right] \left( x\right) =\psi (-\tfrac{b}{2}x^{2})f\left(
x\right) .
\end{equation*}

\item The Weyl element $\mathrm{w}$ acts by discrete Fourier transform 
\begin{equation*}
F\left[ f\right] \left( y\right) =\frac{1}{\sqrt{p}}\sum \limits_{x\in 
\mathbb{F}_{p}}\psi \left( yx\right) f\left( x\right) .
\end{equation*}
\end{itemize}

\bigskip

The above formulas can be easily verified using identity (\ref{Egorov}).
Hence, we conclude that every operator $\rho \left( g\right) ,$ $g\in Sp,$
can be written either in the form $\rho \left( g\right) =M_{_{b}}\circ S_{a}$
or in the form $\rho \left( g\right) =M_{b_{2}}\circ F\circ M_{b_{1}}\circ
S_{a}$. In particular, given a function $f\in 
\mathbb{C}
\left( \mathbb{F}_{p}\right) $, \ applying formula (\ref{relation_eq}) with $%
\phi =\rho \left( s\right) ^{-1}\delta _{1}$ yields \ 
\begin{equation}
\Theta _{T,\phi }\left[ f\right] \left( Ad_{s}^{\vee }\left( \chi \right)
\right) =\frac{1}{p-1}\sum \limits_{a\in \mathbb{F}_{p}^{\times }}\sigma
\left( a\right) \overline{\chi }\left( a\right) \rho \left( s\right)
[f]\left( a\right) ,  \label{fast_eq}
\end{equation}%
for every $\chi \in A^{\vee }$, where $s$ is the specific element that
conjugate $T$ to $A$.

In conclusion, formula (\ref{fast_eq}) implies that $\Theta _{T,\phi }\left[
f\right] $ can be computed by, first, applying the (DFT type) operator $\rho
\left( s\right) $ to $f$ and then applying Mellin transform to the result.
This completes our construction of the Fast Oscillator Transform in the
split case.

\begin{problem}
\label{question}Does there exists a fast algorithm for computing the
oscillator transform associated to a \textbf{non-split} torus?
\end{problem}

\section{Diagonalization of the discrete Fourier transform\label{Diag}}

In this section we apply the previous development in order to exhibit a
canonical basis of eigenvectors for the DFT.

\subsection{Relation between the DFT and the Weil representation}

We will show that the DFT can be naturally identified (up to a normalization
scalar) with an operator $\rho \left( \mathrm{w}\right) $ in the Weil
representation, where $\mathrm{w}$ is an element in a maximal torus $T_{%
\mathrm{w}}\subset Sp$. We take $\mathrm{w}\in Sp$ to be the Weyl element (%
\ref{Weyl}).

\begin{theorem}
\label{DFT_lemma}We have 
\begin{equation*}
F=C\cdot \rho \left( \mathrm{w}\right) ,
\end{equation*}%
where $C=i^{\frac{p-1}{2}}.$

\begin{corollary}
\label{Fq} In case one considers a field extension of the form $\mathbb{F}%
_{q},$ $q=p^{d}$ and the additive character $\psi _{q}:\mathbb{F}%
_{q}\rightarrow 
\mathbb{C}
^{\times }$, given by $\psi _{q}(x)=\psi \left( tr\left( x\right) \right) $.
In this case, the Fourier transform $F$ associated with $\psi _{q}$ and the
operator $\rho \left( \mathrm{w}\right) $ coming from the Weil
representation of $SL_{2}(\mathbb{F}_{q})$, are related by 
\begin{equation*}
F=C\cdot \rho \left( \mathrm{w}\right) ,
\end{equation*}
with $C=(-1)^{d-1}i^{d\frac{p-1}{2}}$.
\end{corollary}
\end{theorem}

For the proofs see Appendix \ref{proofs_appendix}.

\subsection{Diagonalization of the DFT}

Theorem \ref{DFT_lemma} implies that the diagonalization problems of the
operators $F$ and $\rho \left( \mathrm{w}\right) $ are equivalent. The
second problem can be approached using representation theory, which is what
we are going to do next.

Let us denote by $T_{\mathrm{w}}$ the centralizer of $\mathrm{w}$ in $Sp$,
namely $T_{\mathrm{w}}$ consists of all elements $g\in Sp$ such that $g%
\mathrm{w}=\mathrm{w}g$, in particular we have that $\mathrm{w}\in T_{%
\mathrm{w}}$.

\begin{proposition}
\label{fourier_torus_prop}The group $T_{\mathrm{w}}$ is a maximal torus.
Moreover the split type of $T_{\mathrm{w}}$ depends on the prime $p$ in the
following way: $T_{\mathrm{w}}$ is a split torus when $p\equiv 1\left( \func{%
mod}4\right) $ and is a non-split torus when $p\equiv 3\left( \func{mod}%
4\right) $.
\end{proposition}

For a proof see Appendix \ref{proofs_appendix}.

Proposition \ref{fourier_torus_prop} has several consequences. First
consequence is that choosing a unit character vector $\phi _{\chi }\in 
\mathcal{H}_{\chi }$ for every $\chi \in \mathsf{Spec}_{T_{\mathrm{w}%
}}\left( \mathcal{H}\right) $ gives a canonical (up to normalizing unitary
constants) choice of eigenvectors for the DFT \footnote{%
In the case $T_{w}$ is a split torus there is a slight ambiguity in the
choice of a character vector with respect to $\sigma _{T_{w}}$. This
ambiguity can be resolved by further investigation which we will not discuss
here.}. Second, more mysterious consequence is that although the formula of
the DFT is uniform in $p$, its qualitative behavior changes dramatically
between the cases when $p\equiv 1\left( \func{mod}4\right) $ and $p\equiv
3\left( \func{mod}4\right) $. This is manifested in the structure of the
group of symmetries: In the first case, the group of symmetries is a split
torus consisting of $p-1$ elements and in the second case it is a non-split
torus consisting of $p+1$ elements. It also seems that the structure of the
symmetry group is important from the algorithmic perspective, in the case $%
p\equiv 1\left( \func{mod}4\right) $ we built a fast algorithm for computing 
$\Theta =\Theta _{T_{\mathrm{w}}},$ while in the case $p\equiv 3\left( \func{%
mod}4\right) $ the existence of such an algorithm remains open (see Problem %
\ref{question}).

For the convenience of the reader, we enclose in Appendix \ref{EOT} an
explicit formula for $\Theta _{T_{\mathrm{w}}}$ in the case $p\equiv 1\left( 
\func{mod}4\right) .$

\subsection{Multiplicities of eigenvalues of the DFT}

Considering the group $T_{\mathrm{w}}$ we can give a transparent computation
of the eigenvalues multiplicities for the operator $\rho \left( \mathrm{w}%
\right) $. First we note that, since $\mathrm{w}$ is an element of order $4$%
, the eigenvalues of $\rho \left( \mathrm{w}\right) $ lies in the set $%
\left
\{ \pm 1,\pm i\right \} $. For $\lambda \in \left \{ \pm 1,\pm
i\right \} $, let $m_{\lambda }$ denote the multiplicity of the eigenvalue $%
\lambda $. We observe that%
\begin{equation*}
m_{\lambda }=\bigoplus \limits_{\chi \in I_{\lambda }}\dim \mathcal{H}_{\chi
},
\end{equation*}%
where $I_{\lambda }$ consists of all characters $\chi \in \mathsf{Spec}_{T_{%
\mathrm{w}}}\left( \mathcal{H}\right) $ such that $\chi \left( \mathrm{w}%
\right) =\lambda $. The result now follows easily from Theorem \ref{dec_thm}%
, applied to the torus $T_{\mathrm{w}}$. We treat separately the split and
non-split cases.

\begin{itemize}
\item Assume $T_{\mathrm{w}}$ is a split torus, which happens when $%
p=1\left( \func{mod}4\right) $, namely, $p=4l+1$, $l\in 
\mathbb{N}
$. Since $\dim \mathcal{H}_{\chi }=1$ for $\chi \neq \sigma _{T_{\mathrm{w}%
}} $ it follows that $m_{\pm i}=\frac{p-1}{4}=l$. We are left to determine
the values of $m_{\pm 1}$, which depend on whether $\sigma _{T_{\mathrm{w}%
}}\left( \mathrm{w}\right) =\mathrm{w}^{\frac{p-1}{2}}$ is $1$ or $-1$.
Since $\mathrm{w}$ is an element of order $4$ in $T_{\mathrm{w}}$ we get
that 
\begin{equation*}
\sigma _{T_{\mathrm{w}}}\left( \mathrm{w}\right) =\left \{ 
\begin{array}{cc}
1, & p\equiv 1\left( \func{mod}8\right) , \\ 
-1,\text{ \ } & \text{\ }p\equiv 5\left( \func{mod}8\right) ,%
\end{array}%
\right.
\end{equation*}

which implies that when $p\equiv 1\left( \func{mod}8\right) $ then $%
m_{1}=l+1 $ and $m_{-1}=l$ and when $p\equiv 5\left( \func{mod}8\right) $
then $m_{1}=l $ and $m_{-1}=l+1$.

\item Assume $T_{\mathrm{w}}$ is a non-split torus, which happens when $%
p\equiv 3\left( \func{mod}4\right) $, namely, $p=4l+3$, $l\in 
\mathbb{N}
$. Since $\dim \mathcal{H}_{\chi }=1$ for $\chi \neq \sigma _{T_{\mathrm{w}%
}} $ it follows that $m_{\pm i}=\frac{p+1}{4}=l+1$. The values of $m_{\pm 1}$
depend on whether $\sigma _{T_{\mathrm{w}}}\left( \mathrm{w}\right) =\mathrm{%
w}^{\frac{p+1}{2}}$ is $1$ or $-1$. Since $\mathrm{w}$ is an element of
order $4$ in $T_{\mathrm{w}}$ we get that 
\begin{equation*}
\sigma _{T_{\mathrm{w}}}\left( \mathrm{w}\right) =\left \{ 
\begin{array}{cc}
1, & p\equiv 7\left( \func{mod}8\right) , \\ 
-1,\text{ \ } & p\equiv 3\left( \func{mod}8\right) ,%
\end{array}%
\right.
\end{equation*}

which implies that when $p\equiv 7\left( \func{mod}8\right) $ then $m_{1}=l$
and $m_{-1}=l+1$ and when $p\equiv 3\left( \func{mod}8\right) $ then $%
m_{1}=l+1$ and $m_{-1}=l$.
\end{itemize}

Summarizing, the multiplicities of the operator $\rho \left( \mathrm{w}%
\right) $ are 
\begin{equation}
\begin{tabular}{|l|l|l|l|l|}
\hline
& $m_{1}$ & $m_{-1}$ & $m_{i}$ & $m_{-i}$ \\ \hline
$p=8k+1$ & $2k+1$ & $2k$ & $2k$ & $2k$ \\ \hline
$p=8k+3$ & $2k$ & $2k+1$ & $2k+1$ & $2k+1$ \\ \hline
$p=8k+5$ & $2k+1$ & $2k+2$ & $2k+1$ & $2k+1$ \\ \hline
$p=8k+7$ & $2k+2$ & $2k+1$ & $2k+2$ & $2k+2$ \\ \hline
\end{tabular}
\label{table1}
\end{equation}

Considering now the DFT operator $F$. If we denote by $n_{\mu }$, $\mu \in
\left \{ \pm 1,\pm i\right \} $ the multiplicity of the eigenvalue $\mu $ of 
$F $ then the values of $n_{\mu }$ can be deduced from table (\ref{table1})
by invoking the relation $n_{\mu }=m_{\lambda }$ where $\lambda =i^{\frac{p-1%
}{2}}\cdot \mu $ (see Theorem \ref{DFT_lemma}). Summarizing, the
multiplicities of the DFT are 
\begin{equation*}
\begin{tabular}{|l|l|l|l|l|}
\hline
& $n_{1}$ & $n_{-1}$ & $n_{i}$ & $n_{-i}$ \\ \hline
$p=4l+1$ & $l+1$ & $l$ & $l$ & $l$ \\ \hline
$p=4l+3$ & $l+1$ & $l+1$ & $l+1$ & $l$ \\ \hline
\end{tabular}%
\end{equation*}

For a comprehensive treatment of the multiplicity problem from a more
classical point of view see \cite{AT}. \ Other applications of Theorem \ref%
{DFT_lemma} appear in \cite{GHH}.

\bigskip \appendix

\section{Proofs\label{proofs_appendix}}

\subsection{Proof of Theorem \protect \ref{linearization_thm}\label{PL}}

We are left to proof the uniqueness in the case $p\neq 3.$ Since any two
linearizations differ by a character of $SL_{2}(\mathbb{\mathbb{F}}_{p})$,
the uniqueness follows from the fact that $SL_{2}(\mathbb{\mathbb{F}}_{p})$
is a perfect group, namely, it admits no non-trivial characters.

For the sake of completeness let us prove the last assertion

\begin{proposition}
\label{perfect}The group $SL_{2}(\mathbb{F}_{p})$ is perfect.
\end{proposition}

Indeed, let us consider a character $\chi :SL_{2}(\mathbb{F}_{p})\rightarrow 
\mathbb{C}
^{\times }$. On the one hand, considering the pull-back character $\chi
\circ pr$, where $pr:SL_{2}(\mathbb{%
\mathbb{Z}
})\rightarrow SL_{2}(\mathbb{F}_{p})$ is the canonical projection, one
obtains that $\left( \chi \circ pr\right) ^{12}=1$ since the group of
characters of $SL_{2}(\mathbb{%
\mathbb{Z}
})$ is isomorphic to $%
\mathbb{Z}
/12%
\mathbb{Z}
$ \cite{CM}. Now, since $pr$ is a surjection, this implies that $\chi
^{12}=1 $.

On the other hand, the group $SL_{2}(\mathbb{F}_{p})$ is generated by the
unipotent elements 
\begin{equation*}
u_{+}=%
\begin{pmatrix}
1 & 1 \\ 
0 & 1%
\end{pmatrix}%
,\text{ \ }u_{-}=%
\begin{pmatrix}
1 & 0 \\ 
1 & 1%
\end{pmatrix}%
,
\end{equation*}

and since $u_{+}^{p}=u_{-}^{p}=1$, we have $\chi ^{n}=1.$ Hence, $\chi =1$
because $\gcd (12,p)=1.$

This concludes the proof of the theorem.

\subsection{Proof of Theorem \protect \ref{dec_thm}}

Fix a character $\chi :T\rightarrow 
\mathbb{C}
^{\times }$. Let $P_{\chi }$ denote the orthogonal projector on the subspace 
$\mathcal{H}_{\chi }$. We have $\dim \mathcal{H}_{\chi }=Tr\left( P_{\chi
}\right) $. The projector $P_{\chi }$ can be written in terms of the Weil
representation, namely 
\begin{equation*}
P_{\chi }=\frac{1}{\#T}\sum \limits_{g\in T}\overline{\chi }\left( g\right)
\rho \left( g\right) .
\end{equation*}%
Therefore 
\begin{eqnarray*}
Tr\left( P_{\chi }\right) &=&\frac{1}{\#T}\sum \limits_{g\in T}\overline{%
\chi }\left( g\right) Tr\left( \rho \left( g\right) \right) \\
&=&\frac{\dim \mathcal{H}}{\#T}\sum \limits_{g\in T}\overline{\chi }\left(
g\right) K_{g}\left( 0\right) ,
\end{eqnarray*}%
where in the second equality we use the definition of the kernel function $%
K_{g}$. Invoking formula (\ref{inv_formula}) we obtain%
\begin{equation}
Tr\left( P_{\chi }\right) =\frac{1}{\#T}\left( \sigma \left( -1\right)
\left( \sum \limits_{g\in T^{\times }}\overline{\chi }\left( g\right) \sigma
\left( \det \left( \kappa \left( g\right) +I\right) \right) \right) +\dim 
\mathcal{H}\right) ,  \label{basic_formula}
\end{equation}%
where $T^{\times }=T-\{1\}$.

\begin{lemma}
\label{tech_lemma}For every $g\in T^{\times }$, $\sigma \left( \det \left(
\kappa \left( g\right) +I\right) \right) =\sigma _{T}\left( -1\right) \sigma
_{T}\left( g\right) $.
\end{lemma}

Using the above lemma and substituting $\dim \mathcal{H}=p$ in (\ref%
{basic_formula}) we obtain the following basic formula%
\begin{equation}
Tr\left( P_{\chi }\right) =\frac{1}{\#T}\left( \sigma \left( -1\right)
\sigma _{T}\left( -1\right) \left( \sum \limits_{g\in T^{\times }}\overline{%
\chi }\left( g\right) \sigma _{T}\left( g\right) \right) +p\right) .
\label{basic_formula1}
\end{equation}

Next, we analyze (\ref{basic_formula1}) separately according to the split
type of $T$. In our computations we use the simple yet important fact that
the quadratic character $\sigma _{T}$ can be written explicitly as $\sigma
_{T}\left( g\right) =g^{\frac{\#T}{2}}\in \left \{ 1,-1\right \} $.

\begin{itemize}
\item $T$ split. In this case $\#T=p-1$, thus $\sigma _{T}\left( g\right)
=g^{\frac{p-1}{2}}$, which implies that $\sigma \left( -1\right) \sigma
_{T}\left( -1\right) =1$. Consequently 
\begin{eqnarray*}
Tr\left( P_{\chi }\right) &=&\frac{1}{p-1}\left( \left( \sum \limits_{g\in
T^{\times }}\overline{\chi }\left( g\right) \sigma _{T}\left( g\right)
\right) +p\right) \\
&=&\left \{ 
\begin{array}{cc}
\frac{1}{p-1}\left( -1+p\right) & \text{\ }\chi \neq \sigma _{T}, \\ 
\frac{1}{p-1}\left( 2p-2\right) & \chi =\sigma _{T},%
\end{array}%
\right. \\
&=&\left \{ 
\begin{array}{cc}
1 & \text{\ }\chi \neq \sigma _{T}, \\ 
2 & \chi =\sigma _{T}.%
\end{array}%
\right.
\end{eqnarray*}

\item $T$ non-split. In this case $\#T=p+1$, thus $\sigma _{T}\left(
g\right) =g^{\frac{p+1}{2}}$, which implies that $\sigma \left( -1\right)
\sigma _{T}\left( -1\right) =\left( -1\right) ^{\frac{p-1}{2}+\frac{p+1}{2}%
}=-1$. Consequently 
\begin{eqnarray*}
Tr\left( P_{\chi }\right) &=&\frac{1}{p+1}\left( -\left( \sum \limits_{g\in
T^{\times }}\overline{\chi }\left( g\right) \sigma _{T}\left( g\right)
\right) +p\right) \\
&=&\left \{ 
\begin{array}{cc}
\frac{1}{p+1}\left( 1+p\right) & \text{\ }\chi \neq \sigma _{T}, \\ 
\frac{1}{p+1}\left( -p+p\right) & \chi =\sigma _{T},%
\end{array}%
\right. \\
&=&\left \{ 
\begin{array}{cc}
1 & \text{\ }\chi \neq \sigma _{T}, \\ 
0 & \chi =\sigma _{T}.%
\end{array}%
\right.
\end{eqnarray*}
\end{itemize}

This concludes the proof of the theorem.

\subsubsection{Proof of Lemma \protect \ref{tech_lemma}}

Let $U\subset Sp$ be the subset consisting all elements $g\in Sp$ such that $%
g-I$ is invertible. Define $\varepsilon :U\rightarrow 
\mathbb{C}
$, as $\varepsilon \left( g\right) =\sigma \left( \det \left( \kappa \left(
g\right) +1\right) \right) $. We note that $U$ is closed under conjugation
and $\varepsilon $ is conjugation invariant. We treat the split and
non-split cases separately.

\textbf{Assume }$T$\textbf{\ split}. Since $\varepsilon $ is conjugation
invariant, it is enough to show that $\varepsilon \left( g\right) =$ $\sigma
_{T}\left( -1\right) \sigma _{T}\left( g\right) $ when $T$ is the standard
diagonal torus $A$. Let $g=%
\begin{pmatrix}
a & 0 \\ 
0 & a^{-1}%
\end{pmatrix}%
$ $\in A$. We have 
\begin{eqnarray*}
\det \left( \kappa \left( g\right) +1\right) &=&\left( 1+\frac{a+1}{a-1}%
\right) \left( 1-\frac{a+1}{a-1}\right) \\
&=&\frac{-4a}{\left( a-1\right) ^{2}},
\end{eqnarray*}

therefore $\varepsilon \left( g\right) =\sigma \left( -1\right) \sigma
\left( a\right) =\sigma _{T}\left( -1\right) \sigma _{T}\left( g\right) $.
Concluding the argument in this case.

\textbf{Assume }$T$\textbf{\ is non-split.} Again by conjugation invariance,
we can assume that $T$ consists of elements $g\in \mathbb{F}_{p^{2}}^{\times
}$ such that $\det \left( g\right) =g^{p+1}=1$. We have 
\begin{equation*}
\kappa \left( g\right) +1=\frac{g+1}{g-1}+1=\frac{2g}{g-1},
\end{equation*}%
hence 
\begin{equation*}
\det \left( \kappa \left( g\right) +1\right) =\left( \frac{2g}{g-1}\right)
^{p+1}=\frac{4}{\left( g-1\right) ^{p+1}},
\end{equation*}%
and%
\begin{equation*}
\varepsilon \left( g\right) =\sigma \left( \frac{4}{\left( g-1\right) ^{p+1}}%
\right) =\left( \frac{1}{\left( g-1\right) ^{p+1}}\right) ^{\frac{p-1}{2}%
}=\left( \frac{1}{\left( g-1\right) ^{p-1}}\right) ^{\frac{p+1}{2}}.
\end{equation*}%
Explicit computation reveals that $\frac{1}{\left( g-1\right) ^{p-1}}=-g$,
therefore we obtain 
\begin{equation*}
\varepsilon \left( g\right) =\left( -g\right) ^{\frac{p+1}{2}}=\sigma
_{T}\left( -1\right) \sigma _{T}\left( g\right) \text{.}
\end{equation*}%
Concluding the argument in this case and the proof of the lemma.

\subsection{Proof of Lemma \protect \ref{integral_lemma}}

The proof is by direct verification. Write 
\begin{eqnarray*}
\mathcal{M}_{T}\circ m_{T}\left[ \chi \right] &=&\frac{1}{\#T}\sum
\limits_{g\in T}\overline{\chi }\left( g\right) \left \langle f,\rho \left(
g\right) ^{-1}\phi \right \rangle \\
&=&\left \langle f,\frac{1}{\#T}\sum \limits_{g\in T}\chi \left( g\right)
\rho \left( g^{-1}\right) \phi \right \rangle \\
&=&\left \langle f,\left( \frac{1}{\#T}\sum \limits_{g\in T}\overline{\chi }%
\left( g\right) \rho \left( g\right) \right) \phi \right \rangle \\
&=&\left \langle f,P_{\chi }\phi \right \rangle =\left \langle f,\phi _{\chi
}\right \rangle =\Theta _{T,\phi }\left[ f\right] \left( \chi \right) \text{.%
}
\end{eqnarray*}

\subsection{Proof of Lemma \protect \ref{relation_lemma}}

The proof is by direct computation. Let $f\in \mathcal{H}$ and $\chi \in
A^{\vee }$. 
\begin{eqnarray*}
\left( Ad_{s}^{\vee }\right) ^{\ast }\circ \Theta _{T,\phi }\left[ f\right]
\left( \chi \right) &=&\Theta _{T,\phi }\left[ f\right] \left( Ad_{s}^{\vee
}\left( \chi \right) \right) \\
&=&M_{T}\circ m_{T,\phi }\left[ f\right] \left( Ad_{s}^{\vee }\left( \chi
\right) \right) \\
&=&\frac{1}{\#T}\sum \limits_{g\in T}\overline{Ad_{s}^{\vee }\left( \chi
\right) }\left( g\right) \left \langle f,\rho \left( g^{-1}\right) \phi
\right \rangle \\
&=&\frac{1}{\#T}\sum \limits_{g\in T}\overline{\chi }\left( sgs^{-1}\right)
\left \langle f,\rho \left( g^{-1}\right) \phi \right \rangle \\
&=&\frac{1}{\#T}\sum \limits_{g\in A}\overline{\chi }\left( g\right) \left
\langle f,\rho \left( \left( s^{-1}gs\right) ^{-1}\right) \phi \right \rangle
\\
&=&\frac{1}{\#T}\sum \limits_{g\in A}\overline{\chi }\left( g\right) \left
\langle \rho \left( s\right) f,\rho \left( g^{-1}\right) \rho \left(
s\right) \phi \right \rangle \\
&=&\frac{1}{\#A}\sum \limits_{g\in A}\overline{\chi }\left( g\right) \left
\langle \rho \left( s\right) f,\rho \left( g^{-1}\right) \rho \left(
s\right) \phi \right \rangle \\
&=&M_{A}\circ m_{A,\rho \left( s\right) \phi }\left[ \rho \left( s\right) f%
\right] \left( \chi \right) =\Theta _{A,\rho \left( s\right) \phi }\left[
\rho \left( s\right) f\right] \left( \chi \right) .
\end{eqnarray*}

This concludes the proof of the lemma.

\subsection{Proof of Theorem \protect \ref{DFT_lemma}}

The operator $\rho \left( \mathrm{w}\right) $ is characterized up to a
unitary scalar by the identity (see formula (\ref{Egorov})) $\rho \left( 
\mathrm{w}\right) \pi \left( h\right) \rho \left( \mathrm{w}\right)
^{-1}=\pi \left( \mathrm{w}\left( h\right) \right) $ for every $h\in H$.
Explicit computation reveals that for every $h\in H$, $F\circ \pi \left(
h\right) \circ F^{-1}=\pi \left( \mathrm{w}\left( h\right) \right) $, which
implies that $F=C\cdot \rho \left( \mathrm{w}\right) $.

Next, computing the determinant of both sides we get $\det \left( F\right)
=C^{p}\cdot \det \left( \rho \left( \mathrm{w}\right) \right) $.

\begin{claim}
\label{detw}\bigskip We have $\det \left( \rho \left( \mathrm{w}\right)
\right) =1.$
\end{claim}

Since $\det (\rho (\mathrm{w}))=1$, we have 
\begin{equation}
\det \left( F\right) =C^{p}.  \label{detC}
\end{equation}

In addition

\begin{proposition}
\label{detF}We have 
\begin{equation}
\det \left( F\right) =i^{\frac{p(p-1)}{2}}.  \label{det(F)}
\end{equation}
\end{proposition}

From the relations $F^{4}=$ $\rho \left( \mathrm{w}\right) ^{4}=Id_{\mathcal{%
H}}$ we get that $C^{4}=1$. Since $\gcd \left( 4,p\right) =1$ we conclude
from (\ref{detC}) and (\ref{det(F)}) that $C=i^{\frac{p-1}{2}}.$

\subsubsection{Proof of Claim \protect \ref{detw}}

In fact, in case $p\neq 3,$ $\rho $ is unimodular, i.e., $\det \left( \rho
\left( g\right) \right) =1$ for every $g\in SL_{2}(\mathbb{F}_{p})$. Indeed,
the perfectness of $SL_{2}(\mathbb{F}_{p})$ (see Proposition \ref{perfect})
implies that $\det \left( \rho \left( g\right) \right) =1$, for every $g\in
SL_{2}(\mathbb{F}_{p})$, since $\det \circ \rho $ is a character of $SL_{2}(%
\mathbb{F}_{p})$. In order to cover also the case of the field $\mathbb{F}%
_{3}$, we give an alternative argument as follow. Since $\det \circ \rho $
is a character of $SL_{2}(\mathbb{F}_{p}),$ we have $(\det \circ \rho (%
\mathrm{w}))^{p}=1$ (see the argument in \ref{PL})$.$ However, the element $%
\mathrm{w}$ is of order $4$, which implies $\det \left( \rho \left( \mathrm{w%
}\right) \right) ^{4}=1.$ The claim now follows since $\gcd (4,p)=1.$

\subsubsection{Proof of Proposition \protect \ref{detF}}

Consider the matrix of $\sqrt{p}F$. If we write $\psi $ in the form $\psi
\left( x\right) =\zeta ^{x}$, $x=0,..,p-1$ and $\zeta =e^{\frac{2\pi i}{p}}$%
, then the matrix of $F$ takes the form $(\zeta ^{yx}:$ $\ x,y\in \left \{
0,1,..,p-1\right \} )$, hence it is a \textit{Vandermonde matrix.} Applying
the formula (\cite{FH}) for the determinant of the Vandermonde matrix, we get%
\begin{eqnarray}
\det (\sqrt{p}F) &=&\tprod \limits_{0\leq x<y\leq p-1}\left( \zeta
^{y}-\zeta ^{x}\right) =\tprod \limits_{0\leq x<y\leq p-1}(\psi (y)-\psi (x))
\label{CalcDet} \\
&=&\tprod \limits_{0\leq x<y\leq p-1}\psi (\tfrac{x+y}{2})\tprod
\limits_{0\leq x<y\leq p-1}\left( \psi (\tfrac{y-x}{2})-\psi (\tfrac{x-y}{2}%
)\right)  \notag \\
&=&\psi (0)\cdot i^{\frac{p(p-1)}{2}}\cdot 2^{\frac{p(p-1)}{2}}\tprod
\limits_{j=1}^{p-1}(\sin (\tfrac{\pi \cdot j}{p}))^{p-j},  \notag
\end{eqnarray}%
where the equality $\tprod \limits_{0\leq x<y\leq p-1}\psi (\tfrac{x+y}{2}%
)=\psi (0)$ follows from the fact that $\tsum \limits_{0\leq x<y\leq
p-1}(x+y)=\frac{1}{2}\{ \tsum \limits_{x,y\in \mathbb{F}_{p}}(x+y)-\tsum%
\limits_{x=y\in \mathbb{F}_{p}}(x+y)\}=0-0=0.$

Taking the absolute value on both sides of (\ref{CalcDet}), gives us 
\begin{equation*}
p^{\frac{p}{2}}=2^{\frac{p(p-1)}{2}}\tprod \limits_{j=1}^{p-1}(\sin (\tfrac{%
\pi \cdot j}{p}))^{p-j},
\end{equation*}%
hence $\det \left( F\right) =i^{\frac{p(p-1)}{2}}$ as claimed.

\subsection{Proof of Corollary \protect \ref{Fq}}

Taking the trace on both sides of the equation $F=C\cdot \rho (\mathrm{w})$,
using the explicit formula for the character of the Weil representation
obtained from formula (\ref{inv_formula}), and noting that $\sqrt{q}%
Tr(F)=\tsum \limits_{x\in \mathbb{F}_{q}}\psi _{q}(x^{2})$, the claim
follows from the Hasse--Davenport relation \cite{HD}%
\begin{equation*}
-\tsum \limits_{x\in \mathbb{F}_{q}}\psi _{q}(x^{2})=(-\tsum \limits_{x\in 
\mathbb{F}_{p}}\psi (x^{2}))^{d}\text{.}
\end{equation*}

\subsection{Proof of proposition \protect \ref{fourier_torus_prop}}

In order to show that $T_{\mathrm{w}}$ are the rational points of an
algebraic torus, it is enough to show that the characteristic polynomial of $%
\mathrm{w}\in SL_{2}\left( \mathbb{F}_{p}\right) $ has distinct eigenvalues.
We have $\chi _{\mathrm{w}}\left( t\right) =\det (tI-\mathrm{w})=t^{2}+1$
which has roots $\pm \sqrt{-1}$. Looking at $\chi _{\mathrm{w}}\left(
t\right) $ we see that $T_{\mathrm{w}}$ is a split torus when $-1$ admits a
square root in $\mathbb{F}_{p}$, that is when $\sigma \left( -1\right)
=\left( -1\right) ^{\frac{p-1}{2}}=1$ which happens when $p\equiv 1\left( 
\func{mod}4\right) $. Concluding the proof of the proposition.

\section{Explicit formulas for the oscillator transform\label{EOT}}

Looking at formula (\ref{fast_eq}) of the oscillator transform $\Theta _{T_{%
\mathrm{w}}}$, in the case $p\equiv 1$ $(\func{mod}4)$, we see that in order
to have an explicit description we need to describe the operator $\rho (s)$,
where $s\in Sp$ is an element which conjugates the torus $T_{\mathrm{w}}$ to
the standard torus $A$.

It is enough to describe $\rho \left( s\right) $ up to a unitary scalar,
which is what we are going to do. Since, we assume that $p\equiv 1$ $(\func{%
mod}4)$, there exist an element $\epsilon \in \mathbb{F}_{p}^{\times }$,
which satisfies $\epsilon ^{2}=-1.$ Let 
\begin{equation*}
\text{\ }s=%
\begin{pmatrix}
\frac{1}{2} & \frac{\epsilon }{2} \\ 
\epsilon & 1%
\end{pmatrix}%
.
\end{equation*}

Direct verification reveals that $sT_{\mathrm{w}}s^{-1}=A$. Now, the element 
$s$ can be decomposed according to the Bruhat decomposition%
\begin{equation*}
s=%
\begin{pmatrix}
\frac{1}{2} & \frac{\epsilon }{2} \\ 
\epsilon  & 1%
\end{pmatrix}%
=%
\begin{pmatrix}
1 & 0 \\ 
\frac{2}{\epsilon } & 1%
\end{pmatrix}%
\begin{pmatrix}
0 & 1 \\ 
-1 & 0%
\end{pmatrix}%
\begin{pmatrix}
1 & 0 \\ 
\frac{\epsilon }{4} & 1%
\end{pmatrix}%
\begin{pmatrix}
\frac{2}{\epsilon } & 0 \\ 
0 & \frac{\epsilon }{2}%
\end{pmatrix}%
,
\end{equation*}%
which implies, using the explicit formulas which appears in Section \ref%
{fast_subsub}, that 
\begin{equation*}
\rho (s)=C\cdot M_{\frac{2}{\epsilon }}\circ F\circ M_{\frac{\epsilon }{4}%
}\circ S_{\frac{2}{\epsilon }},
\end{equation*}%
where $C$ is some unitary scalar.


\begin{thebibliography}{99}
\bibitem{AT} Auslander L. and Tolimieri R., Is computing with the finite
Fourier transform pure or applied mathematics? \textit{Bull. Amer. Math.
Soc. (N.S.) }\textbf{1} (1979), no. 6, 847-897.

\bibitem{B} Beyl F.R., The Schur multiplicator of $SL(2,%
\mathbb{Z}
/m%
\mathbb{Z}
)$ and the congruence subgroup property \textit{Math. Zeit. 191} (1986).

\bibitem{CM} Coxeter H.S.M. and Moser, W.O.J., Generators and relations for
discrete groups. \textit{Ergebnisse der Mathematik und ihrer Grenzgebiete,
14. Springer-Verlag, Berlin-New York} (1980).

\bibitem{CT} Cooley J.W. and Tukey J.W., An algorithm for the machine
calculation of complex Fourier series. Math. Comput. 19, 297--301 (1965).

\bibitem{DS} Dickinson B.W. and Steiglitz K., Eigenvectors and functions of
the discrete Fourier transform. \textit{IEEE Tans. Acoustics Speech Signal
Proc. 30 no. 1, 25-31} (1982).

\bibitem{FHKMN} Feichtinger H.G., Hazewinkel M., Kaiblinger N., Matusiak E.
and Neuhauser M., Metaplectic operators on $%
\mathbb{C}
^{n}.$ \textit{Q.J. Math. 59 no. 1, 15-28,} (2008).

\bibitem{FH} Fulton W. and Harris J., Representation theory. A first course. 
\textit{Graduate Texts in Mathematics, 129. Readings in Mathematics.
Springer-Verlag, New York} \ (1991).

\bibitem{G} Gr\"{u}nbaum F. A., The eigenvectors of the discrete Fourier
transform. \textit{J. Math. Anal. Appl. 88, 355-363.} (1982).

\bibitem{GH1} Gurevich S. and Hadani R., The geometric Weil representation. 
\textit{Selecta Mathematica, New Series, Birkh\"{a}user Basel} (Accepted:
Dec. 2006).

\bibitem{GH2} Gurevich S. and Hadani R., Quantization of symplectic vector
spaces over finite fields. \textit{arXiv:0705.4556} (2005).

\bibitem{GHH} Gurevich S., Hadani R. and Howe R., Quadratic reciprocity and
sign of Gauss sum via the finite Weil representation. \textit{Submitted }%
(2008).

\bibitem{GHS1} Gurevich S., Hadani R. and Sochen N., The finite harmonic
oscillator and its associated sequences. \textit{Proceedings of the National
Academy of Sciences of the United States of America (PNAS), July 22, 2008,
vol. 105 no. 29 9869-9873} (2008).

\bibitem{GHS2} Gurevich S., Hadani R. and Sochen N., On some deterministic
dictionaries supporting sparsity. \textit{To appear in the special issue on
sparsity (Editors: A. Cohen, R. DeVore, M. Elad, A. Gilbert) of the Journal
of Fourier Analysis and its Applications }(2008).

\bibitem{GHS3} Gurevich S., Hadani R. and Sochen N., The finite harmonic
oscillator and its application to sequences, communication and radar.\textit{%
\ To appear in the IEEE Transaction on Information Theory} (2008).

\bibitem{HD} Hasse H. and Davenport H., Die Nullstelen der
Kongruenzzetafunktionen in gewissen zyklischen Fallen, \textit{J. Reine
Angew. Math. 172 }(1934), 151-182.

\bibitem{M} Mehta M.L., Eigenvalues and Eigenvectors of the finite Fourier
transform. \textit{J. Math. Phys. 28, no. 4, 781-785 (1987).}

\bibitem{V} Vourdas A. Quantum systems with finite Hilbert space: Galois
fields in quantum mechanics. \textit{J.Phys. A 40, no. 33, R285-R331 (2007).}

\bibitem{W} Weil A., Sur certains groupes d'operateurs unitaires. \textit{%
Acta Math. 111} (1964) 143-211.

\bibitem{We} Weyl H., Gruppentheorie und Quantenmechanik. \textit{1st
edition Hirzel, Leipzig} (1928).
\end{thebibliography}
\end{document}